\documentclass[a4paper]{article}
\usepackage{amsmath}
\usepackage{graphicx}
\usepackage{xy}
\usepackage[thmmarks]{ntheorem}

\title{Strategies of Loop Recombination in Ciliates}
\author{Robert Brijder$^1$, Hendrik Jan Hoogeboom$^1$, and Michael Muskulus$^2$\\
\mbox{ }\\
$^1$ Leiden Institute of Advanced Computer Science, Universiteit Leiden,\\
Niels Bohrweg 1, 2333 CA Leiden, The Netherlands,\\
\texttt{rbrijder@liacs.nl}\\
\mbox{ }\\
$^2$ Mathematical Institute, Universiteit Leiden,\\
Niels Bohrweg 1, 2333 CA Leiden, The Netherlands}

\def\blfootnote{\xdef\@thefnmark{}\@footnotetext}
\long\def\symbolfootnote[#1]#2{\begingroup%
\def\thefootnote{\fnsymbol{footnote}}\footnote[#1]{#2}\endgroup}

\xyoption{all}

\theoremstyle{break} \theorembodyfont{\upshape}

\theoremsymbol{}
\newtheorem{Theorem}{Theorem}
\newtheorem{Lemma}[Theorem]{Lemma}
\newtheorem{Corollary}[Theorem]{Corollary}

\theoremsymbol{\rule{1ex}{1ex}}
\newtheorem{Definition}[Theorem]{Definition}

\theoremstyle{nonumberbreak}
\newtheorem{Proof}{Proof}

\theoremsymbol{}
\newtheorem{Example}{Example}

\begin{document}
\newcommand{\rf}{r \! f}
\newcommand{\pset}[1]{{\mathbf #1}}
\newcommand{\merge}{merge}
\newcommand{\RGVertL}[1]{I_{#1}}
\newcommand{\RGVertR}[1]{I'_{#1}}
\newcommand{\dedge}{\ar@{-}}
\newcommand{\redge}{\ar@2{-}}
\newcommand{\redgr}[1]{\mathcal{R}_{#1}}
\newcommand{\rgrem}[2]{\mathcal{R}_{rem_{#2}(#1)}}
\newcommand{\pcgr}{\mathcal{PC}}
\bibliographystyle{plain}

\date{}
\maketitle

\begin{abstract}
Gene assembly in ciliates is an extremely involved DNA
transformation process, which transforms a nucleus, the
micronucleus, to another functionally different nucleus, the
macronucleus. In this paper we characterize which loop recombination
operations (one of the three types of molecular operations that
accomplish gene assembly) can possibly be applied in the
transformation of a given gene from its micronuclear form to its
macronuclear form. We also characterize in which order these loop
recombination operations are applicable. This is done in the
abstract and more general setting of so-called legal strings.
\end{abstract}

\section{Introduction}
Ciliates are a large group of one-cellular organisms having two
functionally different nuclei: the micronucleus and the
macronucleus. An involved DNA transformation process called gene
assembly transforms a micronucleus into a macronucleus. The process
is accomplished using three types of DNA transformations, which
operate on special DNA sequences called pointers. These three types
of operations are called loop recombination, hairpin recombination,
and double-loop recombination. \symbolfootnote[0]{LIACS Technical
Report 2006-01, Universiteit Leiden, January 2006}

For every gene in its micronuclear form, there can be several
sequences of operations to transform this gene to its macronuclear
form. We call such a sequence a strategy. For a given micronuclear
gene strategies may differ in the number of hairpin and double-loop
recombination operations. It has been shown that the number of loop
recombination operations is independent of the chosen strategy
\cite{EPeo01-3,GeneAssemblyBook}, and that this number can be
efficiently calculated \cite{Extended_paper}.

In this paper we further investigate the loop recombination
operation, called the string negative rule in the string pointer
reduction system, a formal model of gene assembly introduced in
\cite{Equiv_String_Graph_1}. We characterize for a given set of
pointers $D$, whether or not there is a strategy that applies loop
recombination operations on exactly these pointers. We show that
this characterization implies an efficient algorithm that determines
this for given $D$. Also, we characterize the order in which the
pointers of $D$ can possibly be applied in strategies. These results
are obtained using the reduction graph, a graph similar to the
breakpoint graph in the theory of sorting by reversal, introduced in
\cite{Extended_paper}.

This paper is organized as follows. In
Section~\ref{sect_basic_notions} we recall basic notions and
terminology concerning mainly strings and graphs, and in
Section~\ref{sect_sprs} we recall a formal model of the gene
assembly process: the string pointer reduction system.
In Section~\ref{sect_red_graph} we recall the notion of reduction
graph and some theorems related to this notion.
In Section~\ref{sect_pc_graph} we define the pointer-component
graph, a graph that depends on the reduction graph, and we discuss a
natural operation on this graph.
In Section~\ref{sect_spanning_pcgraph} we show that spanning trees
of the pointer-component graphs reveal interesting properties
concerning the string negative rule.
Section~\ref{sect_merging_splitting_components} shows that merging
and splitting of vertices in pointer-component graphs relate to the
removal of pointers.
Using the results of Sections~\ref{sect_spanning_pcgraph} and
\ref{sect_merging_splitting_components}, we characterize in
Section~\ref{sect_char_snr} for a given set of pointers $D$, whether
or not there is a strategy that applies string negative rules on
exactly these pointers.
Section~\ref{sect_order_snr} strengthens results of
Section~\ref{sect_char_snr} by also characterizing in which order
the string negative rules can be applied on the pointers.
We conclude this paper with Section~\ref{sect_conclusion}.

\section{Basic Notions and Notation} \label{sect_basic_notions}
In this section we recall some basic notions concerning functions,
strings, and graphs. We do this mainly to fix the basic notation and
terminology.

The \emph{composition} of functions $f: X \rightarrow Y$ and $g: Y
\rightarrow Z$ is the function $g f: X \rightarrow Z$ such that $(g
f) (x) = g(f(x))$ for every $x \in X$. The restriction of $f$ to a
subset $A$ of $X$ is denoted by $f|A$, and for $D \subseteq X$ we
denote by $f[D]$ the set $\{f(x) \mid x \in D\}$.

We will use $\lambda$ to denote the empty string. For strings $u$
and $v$, we say that $v$ is a \emph{substring of $u$} if $u = w_1 v
w_2$, for some strings $w_1$, $w_2$; we also say that $v$
\emph{occurs in $u$}.

For alphabets $\Sigma$ and $\Delta$, a \emph{homomorphism} is a
function $\varphi: \Sigma^* \rightarrow \Delta^*$ such that
$\varphi(xy) = \varphi(x)\varphi(y)$ and for all $x,y \in \Sigma^*$.
Let $\varphi: \Sigma^* \rightarrow \Delta^*$ be a homomorphism. If
there is a $\Gamma \subseteq \Sigma$ such that
$$
\varphi(a) = \begin{cases} a & a \not\in \Gamma \\ \lambda & a \in
\Gamma \end{cases},
$$
then $\varphi$ is denoted by $erase_{\Gamma}$.

We now turn to graphs. A \emph{(undirected) graph} is a tuple $G =
(V,E)$, where $V$ is a finite set and $E \subseteq \{\{x,y\} \mid
x,y \in V\}$. The elements of $V$ are called \emph{vertices} and the
elements of $E$ are called \emph{edges}. We allow $x = y$, and
therefore edges can be of the form $\{x,x\} = \{x\}$ --- an edge of
this form should be seen as an edge connecting $x$ to $x$, i.e., a
`loop' for a vertex.

Isomorphisms between graphs are defined in the usual way. Two graphs
$G = (V,E)$ and $G' = (V',E')$ are \emph{isomorphic}, denoted by $G
\approx G'$, if there is a bijection $\alpha: V \rightarrow V'$ such
that $\{x,y\} \in E$ iff $\{\alpha(x),\alpha(y)\} \in E'$, for all
$x,y \in V$.
A \emph{walk} in a graph $G$ is a string $\pi = e_1 e_2 \cdots e_n$
over $E$ with $n \geq 1$ such that there are vertices $x_1, x_2,
\ldots, x_{n+1}$ where $e_i = \{x_i,x_{i+1}\}$ and $e_{i+1} =
\{x_{i+1},x_{i+2}\}$ for $1 \leq i < n$ (allowing $\{x,x\} = \{x\}$
for all vertices $x$). We then also say that $\pi$ is a \emph{walk
from $x_1$ to $x_{n+1}$} or a \emph{walk between $x_1$ and
$x_{n+1}$}. We say that walk $\pi$ is \emph{simple} if $x_i \not=
x_j$ for $1 \leq i < j \leq n + 1$. A walk from $v$ to $v$ for some
$v \in V$ is called a \emph{cycle}. Note that a loop is a cycle. We
say that $G$ is \emph{acyclic} if there are no cycles in $G$. We say
that $G$ is \emph{connected} if for every two vertices $v_1$ and
$v_2$ of $G$ with $v_1 \not= v_2$, there is a walk from $v_1$ to
$v_2$.
We say that $G$ is a \emph{tree}, if it is a connected acyclic
graph. If we fix a certain vertex of the tree as the \emph{root},
then the usual terminology of trees is used, such as the
\emph{father} of a vertex, and a \emph{child} of a vertex, etc.
Vertex $x$ is \emph{isolated} in $G$ if there is no edge $e$ of $G$
with $x \in e$. The \emph{restriction of $G$ to $E' \subseteq E$},
denoted by $G|_{E'}$, is $(V,E')$.

Graph $G' = (V',E')$ is an \emph{induced subgraph of $G$} if $V'
\subseteq V$ and $E' = E \cap \{\{x,y\} \mid x,y \in V'\}$. We also
say that $G'$ is the \emph{subgraph of $G$ induced by $V'$}. A
subgraph $H$ of $G$ induced by $V_H \subseteq V$ is a
\emph{connected component of $G$} if $H$ is connected, and for every
edge $e \in E$ either $e \subseteq V_H$ or $e \subseteq V \backslash
V_H$.

A \emph{(undirected) multigraph} is a (undirected) graph $G =
(V,E,\epsilon)$, where parallel edges are possible. Therefore, $E$
is a finite set of edges and $\epsilon: E \rightarrow \{\{x,y\} \mid
x,y \in V\}$ is the \emph{endpoint mapping}. Clearly, if $\epsilon$
is injective, then such a multigraph is equivalent to a (undirected)
graph. We let $\Upsilon_1$ denote the set of undirected multigraphs.

A \emph{2-edge coloured graph} is a (undirected) graph $G =
(V,E_1,E_2,f,s,t)$ where $E_1$ and $E_2$ are two finite (not
necessarily disjoint) sets of edges, $s,t \in V$ are two distinct
vertices called the \emph{source vertex} and the \emph{target
vertex} respectively, and there is a vertex labelling function $f: V
\backslash \{s,t\} \rightarrow \Gamma$. The elements of $\Gamma$ are
the \emph{vertex labels}. We use $\Upsilon_2$ to denote the set of
all 2-edge coloured graphs.

Notions such as isomorphisms, walks, connectedness, and trees carry
over to these two types of graphs. For example, for undirected
multigraph $G = (V,E,\epsilon)$ and $E' \subseteq E$, we have
$G|_{E'} = (V,E',\epsilon|E')$. Care must be taken for isomorphisms.
Two 2-edge coloured graphs $G = (V,E_1,E_2,f,s,t)$ and $G' =
(V',E_1',E_2',f',s',t')$ are \emph{isomorphic}, denoted by $G
\approx G'$, if there is a bijection $\alpha: V \rightarrow V'$ such
that $\alpha(s) = s'$, $\alpha(t) = t'$, $f(v) = f'(\alpha(v))$ for
all $v \in V$, and $\{x,y\} \in E_i \mbox{ iff }
\{\alpha(x),\alpha(y)\} \in E'_i$, for all $x,y \in V$, and $i \in
\{1,2\}$.

For 2-edge coloured graphs $G$, we say that a walk $\pi = e_1 e_2
\cdots e_n$ in $G$ is an \emph{alternating walk in $G$} if, for $1
\leq i < n$, both $e_i \in E_1$ and $e_{i+1} \in E_2$ or the other
way around.

\section{String Pointer Reduction System} \label{sect_sprs}
Three (almost) equivalent formal models for gene assembly were
considered in \cite{Equiv_String_Graph_2, Equiv_String_Graph_1,
GeneAssemblyBook}. In this section we briefly recall one of them:
the string pointer reduction system. This is done mainly to fix the
notation and terminology associated with this model. For a detailed
motivation and other results concerning this model we refer to
\cite{GeneAssemblyBook}. We continue to use the string pointer
reduction system in the remainder of this paper.

We fix $\kappa \geq 2$, and define the alphabet $\Delta =
\{2,3,\ldots,\kappa\}$. For $D \subseteq \Delta$, we define $\bar D
= \{ \bar a \mid a \in D \}$ and $\Pi = \Delta \cup \bar \Delta$.
The elements of $\Pi$ will be called \emph{pointers}. We use the
`bar operator' to move from $\Delta$ to $\bar \Delta$ and back from
$\bar \Delta$ to $\Delta$. Hence, for $p \in \Pi$, $\bar {\bar {p}}
= p$. For a string $u = x_1 x_2 \cdots x_n$ with $x_i \in \Pi$, the
\emph{inverse} of $u$ is the string $\bar u = \bar x_n \bar x_{n-1}
\cdots \bar x_1$. For $p \in \Pi$, we define $\pset{p} =
\begin{cases} p & \mbox{if } p \in \Delta \\ \bar{p} & \mbox{if }
p \in \bar{\Delta}
\end{cases}$, i.e., $\pset{p}$ is the `unbarred' variant of $p$. The
\emph{domain} of a string $v \in \Pi^*$ is $dom(v) = \{ \pset{p}
\mid \mbox{$p$ occurs in $v$} \}$. A \emph{legal string} is a string
$u \in \Pi^*$ such that for each $p \in \Pi$ that occurs in $u$, $u$
contains exactly two occurrences from $\{p,\bar p\}$.

\begin{Definition}
Let $u = x_1 x_2 \cdots x_n$ be a legal string with $x_i \in \Pi$
for $1 \leq i \leq n$. For a pointer $p \in \Pi$ such that
$\{x_i,x_j\} \subseteq \{p,\bar p\}$ and $1 \leq i < j \leq n$, the
\emph{p-interval of $u$} is the substring $x_i x_{i+1} \cdots x_j$.
Two distinct pointers $p,q \in \Pi$ \emph{overlap} in $u$ if both
$\pset{q} \in dom(I_p)$ and $\pset{p} \in dom(I_q)$, where $I_p$
($I_q$, resp.) is the $p$-interval ($q$-interval, resp.) of $u$.
\end{Definition}
\begin{Example}
String $u = \bar 4 3 7 \bar 7 \bar 4 3$ is a legal string. However,
$v = 4 2 4$ is not a legal string. Also, $dom(u) = \{3,4,7\}$ and
$\bar u = \bar 3 4 7 \bar 7 \bar 3 4$. The $3$-interval of $u$ is $3
7 \bar 7 \bar 4 3$, and pointers $3$ and $4$ overlap in $u$.
\end{Example}

The string pointer reduction system consists of three types of
reduction rules operating on legal strings. For all $p,q \in \Pi$
with $\pset{p} \not = \pset{q}$:
\begin{itemize}
\item
the \emph{string negative rule} for $p$ is defined by
$\textbf{snr}_{p}(u_1 p p u_2) = u_1 u_2$,
\item
the \emph{string positive rule} for $p$ is defined by
$\textbf{spr}_{p}(u_1 p u_2 \bar p u_3) = u_1 \bar u_2 u_3$,
\item
the \emph{string double rule} for $p,q$ is defined by
$\textbf{sdr}_{p,q}(u_1 p u_2 q u_3 p u_4 q u_5) = u_1 u_4 u_3 u_2
u_5$,
\end{itemize}
where $u_1,u_2,\ldots,u_5$ are arbitrary (possibly empty) strings
over $\Pi$. We also define $Snr = \{ \textbf{snr}_p \mid p \in \Pi
\}$, $Spr = \{ \textbf{spr}_p \mid p \in \Pi \}$ and $Sdr = \{
\textbf{sdr}_{p,q} \mid p,q \in \Pi, \pset{p} \not = \pset{q} \}$ to
be the sets containing all the reduction rules of a specific type.
For a pointer $p$ and a legal string $u$, if both $p$ and $\bar p$
occur in $u$ then we say that both $p$ and $\bar p$ are
\emph{positive} in $u$; if on the other hand only $p$ or only $\bar
p$ occurs in $u$, then both $p$ and $\bar p$ are \emph{negative} in
$u$.

Note that each of these rules is defined only on legal strings that
satisfy the given form. For example, $\textbf{spr}_{\bar 2}$ is
defined on legal string $\bar 2323$, however $\textbf{spr}_{2}$ is
not defined on this legal string. Also note that for every non-empty
legal string there is at least one reduction rule applicable.
Indeed, every non-empty legal string for which no string positive
rule and no string double rule is applicable must have only
non-overlapping negative pointers, thus there is a string negative
rule which is applicable. This is formalized in
Theorem~\ref{th_legal_string_successful}.

\begin{Definition}
The \emph{domain} of a reduction rule $\rho$, denoted by
$dom(\rho)$, equals the set of unbarred variants of the pointers
that the rule is applied to, i.e., $dom(\textbf{snr}_p) =
dom(\textbf{spr}_p) = \{\pset{p}\}$ and $dom(\textbf{sdr}_{p,q}) =
\{\pset{p},\pset{q}\}$ for $p,q \in \Pi$. For a composition $\varphi
= \varphi_n \ \cdots \ \varphi_2 \ \varphi_1$ of reduction rules
$\varphi_1, \varphi_2, \ldots, \varphi_n$, the \emph{domain},
denoted by $dom(\varphi)$, is the union of the domains of its
constituents, i.e., $dom(\varphi) = dom(\varphi_1) \cup
dom(\varphi_2) \cup \cdots \cup dom(\varphi_n)$.
\end{Definition}
\begin{Example}
The domain of $\varphi = \textbf{snr}_2 \ \textbf{spr}_{\bar 4} \
\textbf{sdr}_{7, 5} \ \textbf{snr}_{\bar 9}$ is $dom(\varphi) =
\{2,4,5,7,9\}$.
\end{Example}
\begin{Definition}
Let $S \subseteq \{Snr,Spr,Sdr\}$. Then a composition $\varphi$ of
reduction rules from $S$ is called an \emph{($S$-)reduction}. Let
$u$ be a legal string. We say that $\varphi$ is a \emph{reduction of
$u$}, if $\varphi$ is a reduction and $\varphi$ is applicable to
(defined on) $u$. A \emph{successful reduction $\varphi$ of $u$} is
a reduction of $u$ such that $\varphi(u) = \lambda$. We then also
say that $\varphi$ is \emph{successful for $u$}. We say that $u$ is
\emph{successful in $S$} if there is a successful $S$-reduction of
$u$.
\end{Definition}
Note that if $\varphi$ is a reduction of $u$, then $dom(\varphi) =
dom(u) \backslash dom(\varphi(u))$.
\begin{Example}
Again let $u = \bar 4 3 7 \bar 7 \bar 4 3$. Then $\varphi_1 =
\textbf{sdr}_{\bar 4,3} \ \textbf{spr}_{7}$ is a successful
$\{Spr,Sdr\}$-reduction of $u$. However, both $\varphi_2 =
\textbf{snr}_3 \ \textbf{spr}_{7}$ and $\varphi_3 = \textbf{snr}_8$
are \emph{not} reductions of $u$.
\end{Example}

Since for every (non-empty) legal string there is an applicable
reduction rule, by iterating this argument, we have the following
well known result.
\begin{Theorem} \label{th_legal_string_successful}
For every legal string $u$ there is a successful reduction of $u$.
\end{Theorem}

\section{Reduction Graph} \label{sect_red_graph}
In this section we recall the definition of reduction graph and some
results concerning this graph. First we give the definition of
pointer removal operations on strings, see also
\cite{Extended_paper}.

\begin{Definition} For a subset $D \subseteq \Delta$, the $D$-removal operation,
denoted by $rem_D$, is defined by $rem_D = erase_{D \cup \bar{D}}$.
We also refer to $rem_D$ operations, for all $D \subseteq \Delta$,
as \emph{pointer removal operations}.
\end{Definition}
\begin{Example}
Let $u = 5 4 3 7 2 5 6 2 \bar 7 3 4 6$ be a legal string. Then for
$D = \{4, 6, 7, 9\}$, we have $rem_D(u) = 5 3 2 5 2 3$. In the
remaining examples we will keep using this legal string $u$.
\end{Example}
Below we restate a lemma from \cite{Extended_paper}. The correctness
of this lemma is easy to verify.

\begin{Lemma} \label{ops_appl}
Let $u$ be a legal string and $D \subseteq \Delta$. Let $\varphi$ be
a composition of reduction rules.
\begin{enumerate}
\item
If $\varphi$ is applicable to $rem_D(u)$ and $\varphi$ does not
contain string negative rules, then $\varphi$ is applicable to $u$.
\item
If $\varphi$ is applicable to $u$ and $dom(\varphi) \subseteq dom(u)
\backslash D$, then $\varphi$ is applicable to $rem_D(u)$.
\item
If $\varphi$ is applicable to both $u$ and $rem_D(u)$, then
$\varphi(rem_D(u)) = rem_D(\varphi(u))$.
\end{enumerate}
\end{Lemma}

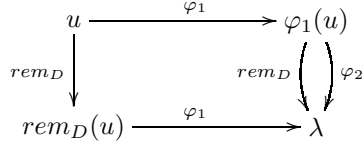
\begin{figure}
$$
\xymatrix{
u \ar[rr]^{\varphi_1} \ar[d]_{rem_D} & & \varphi_1(u) \ar@/^/[d]^{\varphi_2} \ar@/_/[d]_{rem_D}\\
rem_D(u) \ar[rr]^{\varphi_1} & & \lambda
}
$$
\caption{An illustration of Lemma~\ref{ops_appl}.}
\label{ex_lemma_ops_appl}
\end{figure}

Figure~\ref{ex_lemma_ops_appl} illustrates Lemma~\ref{ops_appl} for
the case where there is a successful reduction $\varphi =
\varphi_{2} \ \varphi_{1}$ of $u$, where $\varphi_{1}$ is a
$\{Spr,Sdr\}$-reduction and $\varphi_{2}$ is a $\{Snr\}$-reduction
with $dom(\varphi_{2}) = D$.

We now restate the definition of reduction graph (see
\cite{Extended_paper}) in a less general form. We refer to
\cite{Extended_paper} for a motivation and for more examples and
results concerning this graph. The notion is similar to the
breakpoint graph (or reality-and-desire diagram) known from another
branch of DNA processing theory called sorting by reversal, see e.g.
\cite{SetubalMeidanisBook} and \cite{PevznerBook}.

\begin{Definition}
Let $u = p_1 p_2 \cdots p_n$ with $p_1,\ldots,p_n \in \Pi$ be a
legal string. The \emph{reduction graph of $u$}, denoted by
$\redgr{u}$, is a 2-edge coloured graph $(V,E_1,E_2,f,s,t)$, where
$$
V = \{\RGVertL{1},\RGVertL{2},\ldots,\RGVertL{n}\} \ \cup \
\{\RGVertR{1},\RGVertR{2},\ldots,\RGVertR{n}\} \ \cup \ \{s,t\},
$$
$$
E_{1} = \{e_0, e_1, \ldots, e_{n} \} \mbox{ with }  e_i = \{
\RGVertR{i},\RGVertL{i+1} \} \mbox{ for } 1 < i < n, e_0 =
\{s,\RGVertL{1}\}, e_n = \{ \RGVertR{n}, t \},
$$
\begin{eqnarray*}
E_{2} = & \{ \{\RGVertR{i},\RGVertL{j}\},
\{\RGVertL{i},\RGVertR{j}\} \ | \ i,j \in \{1,2,\ldots,n\}
\mbox{ with } i \not= j \mbox{ and } p_i = p_j \} \ \cup \ \\
& \{ \{\RGVertL{i},\RGVertL{j}\}, \{\RGVertR{i},\RGVertR{j}\} \ | \
i,j \in \{1,2,\ldots,n\} \mbox{ and } p_i = \bar{p}_j \}, \mbox{
and}
\end{eqnarray*}
$$
\mbox{$f(\RGVertL{i}) = f(\RGVertR{i}) = \pset{p_i}$ for $1 \leq i
\leq n$.}
$$
\mbox{  }
\end{Definition}

The edges of $E_1$ are called the \emph{reality edges}, and the
edges of $E_2$ are called the \emph{desire edges}. Notice that for
each $p \in dom(u)$, the reduction graph of $u$ has exactly two
desire edges containing vertices labelled by $p$.

In depictions of reduction graphs, we will represent the vertices
(except for $s$ and $t$) by their labels, because the exact identity
of the vertices is not essential for the problems considered in this
paper. We will also depict reality edges as `double edges' to
distinguish them from the desire edges.

\begin{figure}
$$
\xymatrix{
s \redge[r] & 5 \dedge[r]  & 5 \redge[r] & 6 \dedge[r]  & 6
\redge[r] & t  &
3 \redge[d] \dedge[r] & 3 \redge[d]\\
5 \redge[r] \dedge[d] & 2 \dedge[r]  & 2 \redge[r] & 6 \dedge[d] & 3
\redge[r] \dedge[d] & 4  \dedge[d] &
7  \dedge[d] & 7 \dedge[d] \\
5 \redge[r] & 4 \dedge[r]  & 4 \redge[r] & 6  & 3 \redge[r] & 4
 &
7 \redge[d] & 7 \redge[d]\\
& & & & & & 2  \dedge[r] & 2
}
$$
\caption{The reduction graph of $u$ from the Example.}
\label{ex1_red_graph}
\end{figure}
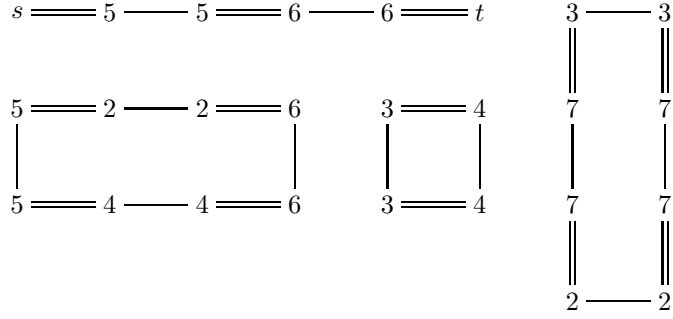

\begin{figure}
$$
\xymatrix{
s \redge[r] & 5 \dedge[r]  & 5 \redge[r] & 6 \dedge[r]
 & 6 \redge[r] & t \\
4 \redge[r] \dedge[d] & 5 \dedge[r]  & 5 \redge[r] & 3 \dedge[d] & 3
\redge[r] \dedge[d] & 4
\dedge[d] \\
4 \redge[r] & 6 \dedge[r]  & 6 \redge[r] & 3 & 3 \redge[r] & 4  }
$$
\caption{The reduction graph $\rgrem{u}{\{2,7\}}$ from the Example.}
\label{ex1_red_graph3}
\end{figure}

\begin{Example}
We continue the example. Reduction graph $\redgr{u}$ is given in
Figure~\ref{ex1_red_graph}, and $\rgrem{u}{\{2,7\}}$ is given in
Figure~\ref{ex1_red_graph3}.
\end{Example}
Each reduction graph has a connected component with a linear
structure containing both the source and the target vertex. This
connected component is called the \emph{linear component} of the
reduction graph. The other connected components are called
\emph{cyclic components} because of their structure.

The definition of reduction functions and the remaining results are
also taken from \cite{Extended_paper}. The $p$-reduction function
removes vertices labelled by $p$ and `contracts' alternating walks
via these vertices into a single edge.

\begin{Definition} \label{def_rf_red}
For each vertex label $p$, we define the \emph{$p$-reduction
function} $\rf_p: \Upsilon_2 \rightarrow \Upsilon_2$, for $G = (V,
E_1, E_2, f, s, t) \in \Upsilon_2$, by
$$
\rf_p(G) = (V', (E_1 \backslash E_{rem}) \cup E_{add}, E_2
\backslash E_{rem}, f|V', s, t),
$$
with
\begin{eqnarray*}
V' & = & \{s,t\} \cup \{ v \in V \backslash \{s,t\} \mid f(v) \not=
p \},
\\
E_{rem} & = & \{ e \in E_1 \cup E_2 \mid f(x) = p \mbox{ for some }
x \in e\}, \mbox{and}
\\
E_{add} & = & \{ \{y_1,y_2\} \mid \mbox{$e_1 e_2 \cdots e_n$ with $n
> 2$ is an alternating walk in $G$}
\\
& & \quad \mbox{with $y_1 \in e_1$, $y_2 \in e_n$, $f(y_1) \not= p
\not= f(y_2)$, and }
\\
& & \quad \mbox{$f(x) = p$ for all $x \in e_i$, $1 < i < n$}\}.
\end{eqnarray*}
\mbox{ }
\end{Definition}

Reduction functions commute under composition. Thus, for a reduction
graph $\rgrem{u}{D}$ and pointers $p$ and $q$, we have
$$
(\rf_q \ \rf_p)(\redgr{u}) = (\rf_p \ \rf_q)(\redgr{u}).
$$
Any reduction can be simulated, on the level of reduction graphs, by
a sequence of reduction functions with the same domain.
\begin{Theorem} \label{th_rf_red}
Let $u$ be a legal string, and let $\varphi$ be a reduction of $u$.
Then
$$
(\rf_{p_n} \ \cdots \ \rf_{p_2} \ \rf_{p_1})(\redgr{u}) \approx
\redgr{\varphi(u)},
$$
where $dom(\varphi) = \{p_1, p_2, \ldots, p_n\}$.
\end{Theorem}

The next lemma is an easy consequence from results in
\cite{Extended_paper}.
\begin{Lemma} \label{lemma_cyclic_comp_2vert}
Let $u$ be a legal string and let $p \in \Pi$. Then $\redgr{u}$ has
a cyclic component $C$ consisting of only vertices labelled by
$\pset{p}$ iff either $pp$ or $\bar p \bar p$ is a substring of $u$.
Moreover, if $C$ exists, then it has exactly two vertices.
\end{Lemma}

One of the motivations for the reduction graph is the easy
determination of the number of string negative rules needed in each
successful reduction \cite{Extended_paper}.
\begin{Theorem} \label{th_cyclic_components1}
Let $N$ be the number of cyclic components in the reduction graph of
legal string $u$. Then every successful reduction of $u$ has exactly
$N$ string negative rules.
\end{Theorem}

\begin{Example}
We continue the example. Since $\redgr{u}$ has three cyclic
components, by Theorem~\ref{th_cyclic_components1}, every successful
reduction $\varphi$ of $u$ has exactly three string negative rules.
For example $\varphi = \textbf{snr}_{6} \ \textbf{snr}_{4} \
\textbf{snr}_{2} \ \textbf{spr}_{\bar{7}} \ \textbf{sdr}_{5,3}$ is a
successful reduction of $u$. Indeed, $\varphi$ has exactly three
string negative rules. Alternatively, $\textbf{snr}_{6} \
\textbf{snr}_{4} \ \textbf{snr}_{3} \ \textbf{spr}_{2} \
\textbf{spr}_{5} \textbf{spr}_{7}$ is also a successful reduction of
$u$, with a different number of ($\textbf{spr}$ and $\textbf{sdr}$)
operations.
\end{Example}

\begin{figure}
$$
\xymatrix{
s \redge[r] & \pset{p}  \dedge[r] & \pset{p} \redge[r] & \pset{q}
 \dedge[r] & \pset{q} \redge[r] & \pset{p}
\dedge[r] & \pset{p} \redge[r] & \pset{q}  \dedge[r] & \pset{q}
\redge[r] & t
}
$$
\caption{The reduction graph of $p q \bar p q$ (and $p q p q$).}
\label{ex_same_red_graph}
\end{figure}

The previous theorem and example should clarify that the reduction
graph reveals crucial properties concerning the string negative
rule. We now further investigate the string negative rule, and show
that many more properties of this rule can be revealed using the
reduction graph.

However, the reduction graph does not seem to be well suited to
prove properties of the string positive rule and string double rule.
If we for example consider legal strings $u = p q \bar p q$ and $v =
p q p q$ for some distinct $p, q \in \Pi$, then $u$ has a unique
successful reduction $\varphi_1 = \textbf{spr}_{\bar q} \
\textbf{spr}_p$ and $v$ has a unique successful reduction $\varphi_2
= \textbf{sdr}_{p,q}$. Thus $u$ must necessarily be reduced by
string positive rules, while $v$ must necessarily be reduced by a
string double rule. However, the reduction graph of $u$ and the
reduction graph of $v$ are isomorphic, as shown in
Figure~\ref{ex_same_red_graph}. Also, whether or not pointers
overlap is not preserved by reduction graphs. For example, the
reduction graphs of legal strings $p q p \bar r q r$ and $p q r \bar
p q r$ for distinct pointers $p$, $q$ and $r$ are isomorphic,
however $p$ and $r$ do not overlap in the first legal string, but
they do overlap in the latter legal string.

The next lemma is an easy consequence of Lemma~\ref{ops_appl} and
Theorem~\ref{th_cyclic_components2}.
\begin{Lemma} \label{lemma_varphi_pc_restr}
Let $u$ be a legal string, and let $D \subseteq dom(u)$. There is a
$\{Spr,Sdr\}$-reduction $\varphi$ of $u$ with $dom(\varphi(u)) = D$
iff $\rgrem{u}{D}$ does not contain cyclic components.
\end{Lemma}
\begin{Proof}
There is a $\{Spr,Sdr\}$-reduction $\varphi$ of $u$ with
$dom(\varphi(u)) = D$ iff there is a successful
$\{Spr,Sdr\}$-reduction of $rem_D(u)$ (by Lemma~\ref{ops_appl}) iff
$\rgrem{u}{D}$ does not contain cyclic components (by
Theorem~\ref{th_cyclic_components1}).
\end{Proof}

We now consider the case where $|D|$ is the number of cyclic
components of $\redgr{u}$.
\begin{Lemma} \label{lemma_varphi_rem_commute}
Let $u$ be a legal string, and let $D \subseteq dom(u)$. There is a
successful reduction $\varphi = \varphi_{2} \ \varphi_{1}$ of $u$,
where $\varphi_{1}$ is a $\{Spr,Sdr\}$-reduction and $\varphi_{2}$
is a $\{Snr\}$-reduction with $dom(\varphi_{2}) = D$ iff
$\rgrem{u}{D}$ and $\redgr{u}$ have $0$ and $|D|$ cyclic components,
respectively.
\end{Lemma}
\begin{Proof}
We first prove the forward implication. By
Lemma~\ref{lemma_varphi_pc_restr}, $\rgrem{u}{D}$ does not contain
cyclic components. By Theorem~\ref{th_cyclic_components1},
$\rgrem{u}{D}$ has $|D|$ cyclic components.

We now prove the reverse implication. By
Lemma~\ref{lemma_varphi_pc_restr}, there is a successful reduction
$\varphi = \varphi_{2} \ \varphi_{1}$ of $u$, where $\varphi_{1}$ is
a $\{Spr,Sdr\}$-reduction and $dom(\varphi_2) = D$. Since
$\redgr{u}$ has $|D|$ cyclic components, by
Theorem~\ref{th_cyclic_components1}, every pointer in $D$ is used in
a string negative rule, and thus $\varphi_2$ is a
$\{Snr\}$-reduction.
\end{Proof}

\section{Pointer-Component Graphs} \label{sect_pc_graph}
If it is clear from the context which legal string $u$ is meant, we
will denote by $\zeta$ the set of connected components of the
reduction graph of $u$. We now define a graph on $\zeta$ that we
will use throughout the rest of this paper. The graph represents how
the labels of a reduction graph are distributed among the connected
components. This graph is particularly useful in determining which
sets $D$ of pointers correspond to strategies that apply loop
recombination operations on exactly the pointers of $D$.
\begin{Definition}
Let $u$ be a legal string. The \emph{pointer-component graph of $u$
(or of $\redgr{u}$)}, denoted by $\pcgr_u$, is an undirected
multigraph $(\zeta, E, \epsilon)$, where $E = dom(u)$ and $\epsilon$
is, for $e \in E$, defined by $\epsilon(e) = \{C \in \zeta \mid C$
$\mbox{contains}$ $\mbox{vertices}$ $\mbox{labelled by } e\}$.
\end{Definition}
Note that for each $e \in dom(u)$, there are exactly two desire
edges connecting vertices labelled by $e$, thus $1 \leq
|\epsilon(e)| \leq 2$, and therefore $\epsilon$ is well defined.

\begin{figure}
$$
\xymatrix{
C_1 \ar@(u,l)@{-}[]_7 \ar@{-}[dr]^2 \\
C_3 \ar@{-}[u]^3 & C_2 \ar@{-}[l]^4 \ar@/^/@{-}[r]^5 & R \ar@/^/@{-}[l]^6 \\
}
$$
\caption{The graph $\pcgr_u$ from the Example.}
\label{ex_cycograph1}
\end{figure}

\begin{Example}
We continue the example. Consider $\redgr{u}$ shown in
Figure~\ref{ex1_red_graph}. Let us define $C_1$ to be the cyclic
component with a vertex labelled by $7$, $C_2$ to be the cyclic
component with a vertex labelled by $5$, $C_3$ to be the third
cyclic component, and $R$ to be the linear component. Then $\zeta =
\{C_1, C_2, C_3, R\}$. The pointer-component graph $\pcgr_u =
(\zeta,dom(u),\epsilon)$ of $u$ is given in
Figure~\ref{ex_cycograph1}.
\end{Example}

We can use the definition of pointer-component graph to reformulate
Theorem~\ref{th_cyclic_components1}.
\begin{Theorem} \label{th_cyclic_components2}
Every successful reduction of a legal string $u$ has exactly
$o(\pcgr_u) - 1$ string negative rules.
\end{Theorem}

For reduction $\varphi$ of a legal string $u$, the difference
between $\redgr{u}$ and $\redgr{\varphi(u)}$ is formulated in
Theorem~\ref{th_rf_red} in terms of reduction functions. We now
reformulate this result for pointer-component graphs. The difference
(up to isomorphism) between the pointer-component graph $PC_1$ of
$\redgr{u}$ and the pointer-component graph $PC_2$ of
$\rf_p(\redgr{u})$ (assuming $\rf_p$ is applicable to $\redgr{u}$)
is as follows: in $PC_2$ edge $p$ is removed and also those vertices
$v$ that become isolated, except when $v$ is the linear component
(since the linear component always contains the source and target
vertex). Since the only legal string $u$ for which the linear
component in $\pcgr_u$ is isolated is the empty string, in this case
we obtain a graph containing only one vertex. This is formalized as
follows. By abuse of notation we will also denote these functions as
reduction functions $\rf_p$.
\begin{Definition} \label{def_rf_pc}
For each edge $p$, we define the \emph{$p$-reduction function}
$\rf_p: \Upsilon_1 \rightarrow \Upsilon_1$, for $G = (V, E,
\epsilon) \in \Upsilon_1$, by
$$
\rf_p(G) = (V', E', \epsilon | E'),
$$
where $E' = E \backslash \{p\}$ and $V' = \{ v \in V \mid v \in
\epsilon(e) \mbox{ for some } e \in E'\}$ if $E' \not= \emptyset$,
and $V' = \{ \emptyset \}$ otherwise.
\end{Definition}
Therefore, these reduction functions correctly simulate (up to
isomorphism) the effect of applications of a reduction functions on
the underlying reduction graph when the reduction functions
correspond to an actual reduction. Note however, when these
reduction functions do not correspond to an actual reduction, the
linear component may become isolated while there are still other
pointers present. Thus in general the reduction functions for
pointer-component graphs do not faithfully simulate the reduction
functions for reduction graphs. As a consequence of
Theorem~\ref{th_rf_red} we obtain now the following result.

\begin{Theorem} \label{th_rf_pc}
Let $u$ be a legal string, and let $\varphi$ be a reduction of $u$.
Then
$$
(\rf_{p_n} \ \cdots \ \rf_{p_2} \ \rf_{p_1})(\pcgr_u) \approx
(\pcgr_{\varphi(u)}),
$$
where $dom(\varphi) = \{p_1, p_2, \ldots, p_n\}.$
\end{Theorem}

\begin{figure}
$$
\xymatrix{
u \ar[rrr]^{\varphi} \ar[d] & & & \varphi(u) \ar[d]\\
\redgr{u} \ar[rrr]^{(\rf_{p_n} \ \cdots \ \rf_{p_1})} \ar[d] & & & \redgr{\varphi(u)} \ar[d]\\
\pcgr_u \ar[rrr]^{(\rf_{p_n} \ \cdots \ \rf_{p_1})} & & &
\pcgr_{\varphi(u)} }
$$
\caption{An illustration of Theorems~\ref{th_rf_red} and
\ref{th_rf_pc}.} \label{ex_th_rf_rem_pc}
\end{figure}
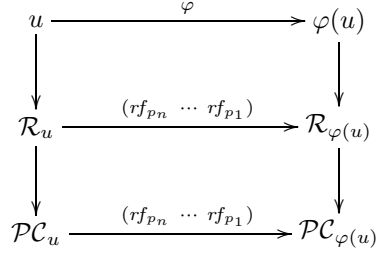
Figure~\ref{ex_th_rf_rem_pc} illustrates Theorems~\ref{th_rf_red}
and \ref{th_rf_pc}.

\begin{figure}
$$
\xymatrix{
C'_1 \ar@(u,l)@{-}[]_7 \ar@{-}[r]^2 & C'_2 \ar@{-}[r]^6 & R' \\
}
$$
\caption{Pointer-component graph $PC_1$ from the Example.}
\label{ex_cycograph2}
\end{figure}
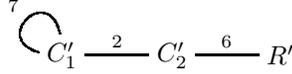

\begin{Example}
We continue the example. We have $(\textbf{snr}_4 \
\textbf{sdr}_{5,3})(u) = 6 2 \bar 7 7 2 6$. The pointer-component
graph $PC_1$ of this legal string is shown in
Figure~\ref{ex_cycograph2}. It is easy to see the graph obtained by
applying $(\rf_{5} \ \rf_{4} \ \rf_{3})$ to $\pcgr_u$
(Figure~\ref{ex_cycograph1}) is isomorphic to $PC_1$.
\end{Example}

\section{Spanning Trees in Pointer-Component Graphs}
\label{sect_spanning_pcgraph} In this section we consider spanning
trees in pointer-component graphs, and we show that there is an
intimate connection between these trees and applicable strategies of
string negative rules. The $\textbf{snr}$ rules in a reduction can
be `postponed' without affecting the applicability. Thus we can
separate each reduction into a sequence without $\textbf{snr}$
rules, and a tail of $\textbf{snr}$ rules. We often use this `normal
form' for notational convenience. First, we characterize
$\{Spr,Sdr\}$-reductions in terms of pointer-component graphs.
\begin{Theorem} \label{th_red_spr_sdr}
Let $\varphi$ be a reduction of legal string $u$, and let $D =
dom(\varphi(u))$. Then $\varphi$ is a $\{Spr,Sdr\}$-reduction of $u$
iff $\pcgr_{\varphi(u)} \approx \pcgr_u|_{D}$.
\end{Theorem}
\begin{Proof}
We first prove the forward implication. By
Theorem~\ref{th_legal_string_successful}, there is a successful
reduction $\varphi'$ of $\varphi(u)$. By
Theorem~\ref{th_cyclic_components2}, $(\varphi' \ \varphi)$ has
$o(\pcgr_u)-1$ string negative rules. Since $\varphi$ is a
$\{Spr,Sdr\}$-reduction of $u$, $\varphi'$ also has $o(\pcgr_u)-1$
string negative rules. Since $\varphi'$ is a successful reduction of
$\varphi(u)$, by Theorem~\ref{th_cyclic_components2},
$\pcgr_{\varphi(u)}$ has the same number of vertices as $\pcgr_u$.
Therefore, by the definition of reduction function and
Theorem~\ref{th_rf_pc}, $\pcgr_{\varphi(u)} \approx \pcgr_u|_{D}$.

We now prove the reverse implication. Let $\varphi'$ be a successful
reduction of $\varphi(u)$. If $\pcgr_{\varphi(u)} \approx
\pcgr_u|_{D}$, then $\pcgr_{\varphi(u)}$ has the same number of
vertices as $\pcgr_u$. Then, by Theorem~\ref{th_cyclic_components2},
$(\varphi' \ \varphi)$ has the same number of string negative rules
as $\varphi$, therefore $\varphi$ is a $\{Spr,Sdr\}$-reduction of
$u$.
\end{Proof}

It will be useful to separate loops from other edges in
pointer-component graphs.
\begin{Definition}
Let $u$ be a legal string and let $\pcgr_u = (V,E,\epsilon)$. We
define $snrdom(u) = \{e \in E \mid |\epsilon(e)| = 2\}$.
\end{Definition}
Thus, $snrdom(u)$ is the set of vertex labels $p$ for which there
are vertices labelled by $p$ in \emph{different} connected
components of $\redgr{u}$.

\begin{Example}
We continue the example. We have $snrdom(u) = \{2,3,4,5,6\}$, and
$dom(u)$ $\backslash$ $snrdom(u) = \{7\}$. Indeed, the only loop in
Figure~\ref{ex_cycograph1} is $7$, indicating that this pointer
occurs only in one connected component of $\redgr{u}$.
\end{Example}
The following corollary to Theorem~\ref{th_rf_pc} observes that an
edge in $dom(\varphi(u))$ is a loop in $\pcgr_{\varphi(u)}$ iff it
is a loop in $\pcgr_u$.

\begin{Corollary} \label{snrdom_cor}
Let $u$ be a legal string and $\varphi$ a reduction of $u$. Then
$snrdom(\varphi(u)) = dom(\varphi(u)) \cap snrdom(u) = snrdom(u)
\backslash dom(\varphi)$.
\end{Corollary}

The examples so far have shown connected pointer-component graphs.
We now prove that these graphs are always connected.
\begin{Theorem} \label{pcgr_connected}
The pointer-component graph of any legal string is connected.
\end{Theorem}
\begin{Proof}
Let $\varphi$ be a successful reduction of a legal string $u$
($\varphi$ exists by Theorem~\ref{th_legal_string_successful}).
Assume that $\pcgr_u$ is not connected. Since $\pcgr_{\lambda}$ is
connected, we have by Theorem~\ref{th_rf_pc} $\varphi = \varphi_2
\rho \varphi_1$ for some reduction rule $\rho$, where
$\pcgr_{\varphi_1(u)}$ is not connected, but $\pcgr_{\rho
\varphi_1(u)}$ is. By Theorem~\ref{th_red_spr_sdr}, $\rho$ cannot be
a string double rule or a string positive rule, and therefore $\rho
= \textbf{snr}_{p}$ for some $p \in \Pi$. Since
$\pcgr_{\varphi_1(u)}$ is not connected, but $\pcgr_{\rho
\varphi_1(u)}$ is, by the definition of reduction function, there is
a connected component of $\pcgr_{\varphi_1(u)}$ containing only the
edge $p$. We consider two cases: $\pset{p} \in snrdom(u)$ and
$\pset{p} \in dom(u) \backslash snrdom(u)$. If $\pset{p} \in
snrdom(u)$, then $\redgr{\rho \varphi_1(u)}$ would have two
connected components less than $\redgr{\varphi_1(u)}$ --- a
contradiction by Theorem~\ref{th_cyclic_components2}. If $\pset{p}
\in dom(u) \backslash snrdom(u)$, then $\pcgr_{\varphi_1(u)}$ has a
connected component containing only a vertex and a loop --- a
contradiction by Lemma~\ref{lemma_cyclic_comp_2vert}. Thus in both
cases we have a contradiction, and therefore $\pcgr_u$ is connected.
\end{Proof}

The next theorem characterizes successfulness in $\{Snr\}$ using
spanning trees.
\begin{Theorem} \label{th_spanning_tree_snr successful}
Let $u$ be a legal string. Then $u$ is successful in $\{Snr\}$ iff
$\pcgr_u$ is a tree.
\end{Theorem}
\begin{Proof}
If $u$ is successful in $\{Snr\}$, then, by
Theorem~\ref{th_cyclic_components2}, $\pcgr_u$ has $|\zeta| - 1$
edges. By Theorem~\ref{pcgr_connected} it follows that $\pcgr_u$ is
a tree.

If $\pcgr_u$ is a tree, then $\pcgr_u$ has $|\zeta| - 1$ edges.
Since the number of edges is $|dom(u)|$, we have $|dom(u)| = |\zeta|
- 1$, and by Theorem~\ref{th_cyclic_components2} every $p \in
dom(u)$ is used in a string negative rule, and thus $u$ is
successful in $\{Snr\}$.
\end{Proof}

It turns out that the pointers on which string negative rules are
applied in a successful reduction of $u$ form a spanning tree of
$\pcgr_u$.
\begin{Theorem} \label{th_snr1_if}
Let $u$ be a legal string, and let $D \subseteq dom(u)$. If there is
a successful reduction $\varphi = \varphi_{2} \ \varphi_{1}$ of $u$,
where $\varphi_{1}$ is a $\{Spr,Sdr\}$-reduction and $\varphi_{2}$
is a $\{Snr\}$-reduction with $dom(\varphi_{2}) = D$, then
$\pcgr_u|_{D}$ is a tree.
\end{Theorem}
\begin{Proof}
By Theorem~\ref{th_spanning_tree_snr successful},
$\pcgr_{\varphi_1(u)}$ is a tree. By Theorem~\ref{th_red_spr_sdr}
$\pcgr_{\varphi_1(u)} \approx \pcgr_u|_{D}$.
\end{Proof}

\begin{Example}
We continue the example. We saw that $\varphi = \textbf{snr}_{6} \
\textbf{snr}_{4} \ \textbf{snr}_{2} \ \textbf{spr}_{\bar{7}} \
\textbf{sdr}_{5,3}$ is a successful reduction of $u$. By
Theorem~\ref{th_snr1_if}, $\pcgr_u|_{\{2,4,6\}}$ is a tree. This is
clear from Figure~\ref{ex_cycograph1} where $\pcgr_u$ is depicted.
\end{Example}

In the next few sections we prove the reverse implication of the
previous theorem. This will require considerably more effort than
the forward implication. The reason for this is that it is not
obvious that when $\pcgr_u|_{D}$ is a tree, there is a reduction
$\varphi_1$ of $u$ such that $D = dom(\varphi_1(u))$. We will use
the pointer removal operation to prove this.

First, we consider a special case of the previous theorem. Since a
loop can never be part of a tree, we have the following corollary to
Theorem~\ref{th_snr1_if}.
\begin{Corollary} \label{snr_cor}
Let $u$ be a legal string and let $p \in dom(u)$. If
$\textbf{snr}_p$ or $\textbf{snr}_{\bar p}$ is in a (successful)
reduction of $u$, then $p \in snrdom(u)$.
\end{Corollary}

\begin{Example}
We continue the example. Since $\varphi = \textbf{snr}_{6} \
\textbf{snr}_{4} \ \textbf{snr}_{2} \ \textbf{spr}_{\bar{7}} \
\textbf{sdr}_{5,3}$ is a successful reduction of $u$, we have $2,4,6
\in snrdom(u)$.
\end{Example}
We show in Theorem~\ref{snr_th} that the reverse implication of
Corollary~\ref{snr_cor} also holds. Hence, the name $snrdom(u)$ is
explained: the pointers $p \in snrdom(u)$ are exactly the pointers
for which $\textbf{snr}_p$ or $\textbf{snr}_{\bar p}$ can occur in a
(successful) reduction of $u$.

\section{Merging and Splitting Components}
\label{sect_merging_splitting_components} In this section we
consider the effect of pointer removal operations on
pointer-component graphs. It turns out that these operations
correspond to the merging and splitting of connected components of
the underlying reduction graph. First, we formally introduce the
merging operation.

\begin{Definition}
For each edge $p$, the \emph{$p$-merge rule}, denoted by $\merge_p$,
is a rule applicable to (defined on) $G = (V, E, \epsilon) \in
\Upsilon_1$ with $p \in E$ and $|\epsilon(p)| = 2$. It is defined by
$$
\merge_p(G) = (V', E', \epsilon'),
$$
where $E' = E \backslash \{p\}$, $V' = \left( V \backslash
\epsilon(p) \right) \cup \{v'\}$ with $\{v'\} \cap V = \emptyset$,
and $\epsilon'(e) = \{h(v_1),h(v_2)\}$ iff $\epsilon(e) =
\{v_1,v_2\}$ where $h(v) = v'$ if $v \in \epsilon(p)$, otherwise it
is the identity.
\end{Definition}

Again, we allow both $v_1 = v_2$ and $h(v_1) = h(v_2)$ in the
previous definition. Intuitively, the $p$-merge rule `merges' the
two endpoints of edge $p$ into one vertex, and therefore the
resulting graph has exactly one vertex less than the original graph.
Note that $p$-merge rules commute under composition. Thus, if
$(\merge_q \ \merge_p)$ is applicable to $G$, then
$$
(\merge_q \ \merge_p)(G) = (\merge_p \ \merge_q)(G).
$$

\begin{Theorem} \label{th_merge_applicable}
Let $G = (V, E, \epsilon) \in \Upsilon_1$, and let $D = \{p_1,
\ldots, p_n\} \subseteq E$. Then $(\merge_{p_n}$ $\cdots \
\merge_{p_1})$ is applicable to $G$ iff $G|_D$ is acyclic.
\end{Theorem}
\begin{Proof}
$(\merge_{p_n} \ \cdots \ \merge_{p_1})$ is applicable on $G$ iff
for all $p_i$ ($1 \leq i \leq n$), $\epsilon(p_i) \not\subseteq
\epsilon[\{p_1,\ldots,p_{i-1}\}]$ and $|\epsilon(p_i)| = 2$ iff
$G|_D$ is acyclic.
\end{Proof}

One of the most surprising aspects of this paper is that the pointer
removal operation is crucial in the proofs of the main results. The
next theorem compares $\pcgr_u$ with $\pcgr_{rem_{\{p\}}(u)}$ for a
legal string $u$ and $p \in dom(u)$. We distinguish three cases:
either the number of vertices of $\pcgr_{rem_{\{p\}}(u)}$ is one
less, is equal, or is one more than the number of vertices of
$\pcgr_u$. The proof of this theorem shows that the first case
corresponds to merging two connected components of $\redgr{u}$ into
one connected component, and the last case corresponds to splitting
one connected component of $\redgr{u}$ into two connected
components.
\begin{Theorem} \label{th_merge_rem}
Let $u$ be a legal string.
\begin{itemize}
\item
If $p \in snrdom(u)$, then $\pcgr_{rem_{\{p\}}(u)} \approx
merge_p(\pcgr_u)$
\\
(and therefore $o(\pcgr_{rem_{\{p\}}(u)}) = o(\pcgr_u) - 1$).
\item
If $p \in dom(u) \backslash snrdom(u)$, then $o(\pcgr_u) \leq
o(\pcgr_{rem_{\{p\}}(u)}) \leq o(\pcgr_u)+1$.
\end{itemize}
\end{Theorem}
\begin{Proof}
We first prove the $p \in snrdom(u)$ case of the theorem. Then the
two desire edges with vertices labelled by $p$ belong to different
connected components of $\redgr{u}$. We distinguish two cases:
whether or not there are cyclic components consisting of only
vertices labelled by $p$.

If there is cyclic component consisting of only vertices labelled by
$p$, then by Lemma~\ref{lemma_cyclic_comp_2vert}, $pp$ or $\bar p
\bar p$ are substrings of $u$, and $\redgr{u}$ is
$$
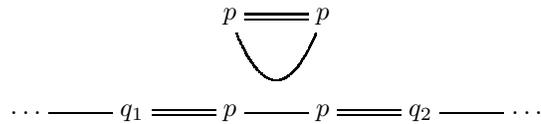
\xymatrix{
& & p \redge[r] \dedge@/_2.0pc/[r] & p  \\
\ldots \dedge[r] &
q_1 \redge[r] &
p  \dedge[r] & p \redge[r] &
q_2  \dedge[r] &
\ldots
}
$$
where we omitted the parts of the graph that are the same compared
to $\rgrem{u}{\{p\}}$. Now, $\rgrem{u}{\{p\}}$ is
$$
\xymatrix{
\ldots \dedge[r] &
q_1 \redge[r] &
q_2  \dedge[r] &
\ldots
}
$$
Therefore $\pcgr_{rem_{\{p\}}(u)}$ can be obtained (up to
isomorphism) from $\pcgr_u$ by applying the $merge_p$ operation.

Now assume that there are no cyclic components consisting of only
vertices labelled by $p$. Then, $\redgr{u}$ is
$$
\xymatrix{
\ldots \dedge[r] &
q_1 \redge[r] &
p  \dedge[r] & p \redge[r] &
q_2  \dedge[r] &
\ldots \\
\ldots \dedge[r] &
q_3 \redge[r] &
p  \dedge[r] & p \redge[r] &
q_4  \dedge[r] &
\ldots
}
$$
where we again omitted the parts of the graph that are the same
compared to $\rgrem{u}{\{p\}}$. Now, depending on the positions of
$q_1, \ldots, q_4$ relative to $p$ in $u$ and on whether $p$ is
positive or negative in $u$, $\rgrem{u}{\{p\}}$ is either
$$
\xymatrix{
\ldots \dedge[r] &
q_1 \redge[r] &
q_4  \dedge[r] &
\ldots \\
\ldots \dedge[r] &
q_3 \redge[r] &
q_2  \dedge[r] &
\ldots
}
$$
or
$$
\xymatrix{
\ldots \dedge[r] &
q_1 \redge[r] &
q_3  \dedge[r] &
\ldots \\
\ldots \dedge[r] &
q_4 \redge[r] &
q_2  \dedge[r] &
\ldots
}
$$
Since $q_1$ and $q_2$ remain part of the same connected component
(the same holds for $q_3$ and $q_4$), the two connected components
are merged, and thus $\pcgr_{rem_{\{p\}}(u)}$ can be obtained (up to
isomorphism) from $\pcgr_u$ by applying the $merge_p$ operation.

We now prove the $p \in dom(u) \backslash snrdom(u)$ case. Then the
two desire edges with vertices labelled by $p$ belong to the same
connected component of $\redgr{u}$. By
Lemma~\ref{lemma_cyclic_comp_2vert}, there are no cyclic components
consisting of four vertices which are all labelled by $p$. We can
distinguish two cases: whether or not there is a reality edge $e$
connecting two vertices labelled by $p$. If there is such an reality
edge $e$ than $\redgr{u}$ is
$$
\xymatrix @=20pt{
\ldots \dedge[r] &
q_1 \redge[r] &
p  \dedge[r] & p \redge[r] &
p  \dedge[r] & p \redge[r] &
q_4  \dedge[r] &
\ldots
}
$$
where we again omitted the parts of the graph that are the same
compared to $\rgrem{u}{\{p\}}$. Thus occurs precisely when $\bar p
p$ or $p \bar p$ is a substring of $u$. Now, $\rgrem{u}{\{p\}}$ is
$$
\xymatrix{
\ldots \dedge[r] &
q_1 \redge[r] &
q_4  \dedge[r] &
\ldots
}
$$
Therefore, $\rgrem{u}{\{p\}}$ has $N$ cyclic components.

If there is no such a reality edge $e$, then $\redgr{u}$ is
$$
\xymatrix @=20pt{
\ldots \dedge[r] &
q_1 \redge[r] &
p  \dedge[r] & p \redge[r] &
q_2  \dedge[r] &
L
\dedge[r] &
q_3 \redge[r] &
p  \dedge[r] & p \redge[r] &
q_4  \dedge[r] &
\ldots
}
$$
where $L$ represents some (possibly empty) `linear subgraph' of
$\redgr{u}$, and where we again omitted the parts of the graph that
are the same compared to $\rgrem{u}{\{p\}}$. Now, $\rgrem{u}{\{p\}}$
is either
$$
\xymatrix @=20pt{
\ldots \dedge[r] &
q_4 \redge[r] &
q_2  \dedge[r] &
L
\dedge[r] &
q_3 \redge[r] &
q_1  \dedge[r] &
\ldots
}
$$
or
$$
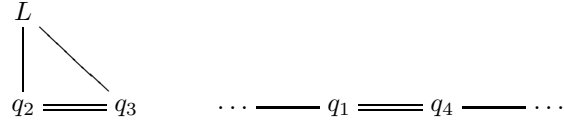
\xymatrix{
& L \dedge[d] \dedge[dr]
\\
& q_2 \redge[r] &
q_3  &
\ldots \dedge[r] &
q_1 \redge[r] &
q_4  \dedge[r] &
\ldots
}
$$
Therefore, $\rgrem{u}{\{p\}}$ has either $N$ cyclic components
(corresponding with the first case) or $N+1$ cyclic components
(corresponding with the second case).
\end{Proof}

\begin{figure}
$$
\xymatrix{
s \redge[r] & 5 \dedge[r] & 5 \redge[r] & 6 \dedge[r]
 & 6 \redge[r] & t \\
4 \redge[r] \dedge[d] & 5 \dedge[r] & 5 \redge[r] & 7 \dedge[r] & 7
\redge[r] & 3  \dedge[d] & 3 \redge[r] \dedge[d] & 4
\dedge[d] \\
4 \redge[r] & 6 \dedge[r] & 6 \redge[r] & 7 \dedge[r] & 7 \redge[r]
& 3  & 3 \redge[r] & 4  }
$$
\caption{Reduction graph $\rgrem{u}{\{2\}}$ from the Example.}
\label{ex1_red_graph2}
\end{figure}

\begin{figure}
$$
\xymatrix{
C_3 \ar@/^/@{-}[r]^3 & C' \ar@(ul,ur)@{-}[]^7 \ar@/^/@{-}[l]^4 \ar@/^/@{-}[r]^5 & R \ar@/^/@{-}[l]^6 \\
}
$$
\caption{$\pcgr_{rem_{\{2\}}(u)}$ from the Example.}
\label{ex_cycograph3}
\end{figure}

\begin{Example}
We continue the example. By Theorem~\ref{th_merge_rem}, we know from
Figure~\ref{ex_cycograph1} that $\pcgr_{rem_{\{2\}}(u)} \approx
merge_2(\pcgr_u)$. Indeed, this is transparent from
Figures~\ref{ex_cycograph1}, \ref{ex1_red_graph2} and
\ref{ex_cycograph3}, where $\pcgr_u$, $\rgrem{u}{\{2\}}$, and
$\pcgr_{rem_{\{2\}}(u)}$ are depicted, respectively.

Again by Theorem~\ref{th_merge_rem}, we know from
Figure~\ref{ex1_red_graph2} that $\rgrem{u}{\{2,7\}}$ has two or
three cyclic components. Indeed, this is transparent from
Figure~\ref{ex1_red_graph3}, where $\rgrem{u}{\{2,7\}}$ is depicted.
\end{Example}

Note that by the definition of $merge_p$, $merge_p$ is applicable to
$\pcgr_u$ precisely when $p \in snrdom(u)$. Therefore, by
Theorems~\ref{th_merge_applicable} and \ref{th_merge_rem}, we have
the following corollary.
\begin{Corollary} \label{cor_merge_rem_it}
Let $u$ be a legal string, and let $D \subseteq dom(u)$. If
$\pcgr_u|_D$ is acyclic, then
$$
\pcgr_{rem_{D}(u)} \approx (\merge_{p_n} \ \cdots \
\merge_{p_1})(\pcgr_u),
$$
where $D = \{p_1, \ldots, p_n\}$.
\end{Corollary}

\section{Applicability of the String Negative Rule}
\label{sect_char_snr} In this section we characterize for a given
set of pointers $D$, whether or not there is a (successful) strategy
that applies string negative rules on exactly these pointers. First
we will prove the following result which depends heavily on the
results of the previous section. The forward implication of the
result observes that by removing pointers from $u$ that form a
spanning tree in $\pcgr_u$ we obtain a legal string $u'$ for which
the reduction graph does not have cyclic components.

\begin{Lemma} \label{lemma_counting_form}
Let $u$ be a legal string, and let $D \subseteq dom(u)$. Then
$\pcgr_u|_{D}$ is a tree iff $\rgrem{u}{D}$ and $\redgr{u}$ have $0$
and $|D|$ cyclic components, respectively.
\end{Lemma}
\begin{Proof}
We first prove the forward implication. Let $\pcgr_u|_{D}$ be a
tree. By Corollary~\ref{cor_merge_rem_it}, $\pcgr_{rem_D(u)}$
contains a single vertex. Thus $\rgrem{u}{D}$ has no cyclic
components. Since $\pcgr_u|_{D}$ is a tree, we have $|D| = |\zeta| -
1$.

We now prove the reverse implication. Let $\rgrem{u}{D}$ not contain
cyclic components and $|D| = |\zeta| - 1$. By
Theorem~\ref{th_merge_rem} we see that the removal of each pointer
$p$ in $D$ corresponds to a $\merge_p$ operation, otherwise
$\rgrem{u}{D}$ would contain cyclic components. Therefore,
$(\merge_{p_n}$ $\cdots \ \merge_{p_1})$ is applicable to $\pcgr_u$
with $D = \{p_1,\ldots,p_n\}$. Therefore, by
Theorem~\ref{th_merge_applicable}, $\pcgr_u|_{D}$ is acyclic. Again
since $|D| = |\zeta| - 1$, it is a tree.
\end{Proof}

\begin{figure}
$$
\xymatrix{
C_3 \ar@{-}[r]^3 & C_1 \ar@{-}[r]^2 & C_2 \ar@{-}[r]^5 & R
}
$$
\caption{A subgraph of the pointer-component graph from the
Example.} \label{ex1_subgraph1}
\end{figure}

\begin{figure}
$$
\xymatrix{
C_1 \ar@{-}[dr]^2 \\
C_3 \ar@{-}[u]^3 & C_2 \ar@{-}[l]^4 & R\\
}
$$
\caption{A subgraph of the pointer-component graph from the
Example.} \label{ex1_subgraph2}
\end{figure}

\begin{figure}
$$
\xymatrix{
s \redge[r] & 4 \dedge[r] & 4 \redge[r] & 6 \dedge[r]
 & 6 \redge[r] & 7 \dedge[r] & 7
\redge[d]
\\
t \redge[r] & 6 \dedge[r]  & 6 \redge[r] & 7 \dedge[r] & 7 \redge[r]
& 4 \dedge[r] & 4
}
$$
\caption{The reduction graph $\rgrem{u}{D_1}$ from the Example.}
\label{ex_red_graph_D1}
\end{figure}

\begin{figure}
$$
\xymatrix{
s \redge[r] & 5 \dedge[r]  & 5 \redge[r] & 6 \dedge[r]  & 6
\redge[r] & t
\\
& 5 \redge[r] \dedge[d] & 7 \dedge[r] & 7 \redge[r] & 6 \dedge[d]
\\
& 5 \redge[r] & 7 \dedge[r] & 7 \redge[r] & 6
\\
}
$$
\caption{The reduction graph $\rgrem{u}{D_2}$ from the Example.}
\label{ex_red_graph_D2}
\end{figure}

\begin{Example}
We continue the previous example. Let $D_1 = \{2,3,5\}$ and $D_2 =
\{2,3,4\}$. Then $\pcgr_u|_{D_1}$ ($\pcgr_u|_{D_2}$, resp.) is given
in Figure~\ref{ex1_subgraph1} (Figure~\ref{ex1_subgraph2}, resp.).
Notice that $|D_1| = |D_2| = |\zeta| - 1$. Since $\pcgr_u|_{D_1}$ is
a tree and $\pcgr_u|_{D_2}$ is not a tree, by
Lemma~\ref{lemma_counting_form}, it follows that $\rgrem{u}{D_1}$
does not have cyclic components and that $\rgrem{u}{D_2}$ does have
at least one cyclic component. This is illustrated in
Figures~\ref{ex_red_graph_D1} and \ref{ex_red_graph_D2}, where
$\rgrem{u}{D_1}$ and $\rgrem{u}{D_2}$ are depicted respectively.
\end{Example}

The next theorem is one of the main results of this paper. It
improves Theorem~\ref{th_snr1_if} by characterizing exactly which
string negative rules can be applied together in a successful
reduction of a given legal string.

In this theorem we require that the string negative rules of
$\varphi$ are applied last. Recall that this is only a notational
convenience since for every successful reduction of a legal string,
we can postpone the application string negative rules to obtain a
successful reduction of the given form.
\begin{Theorem} \label{th_snr1}
Let $u$ be a legal string, and let $D \subseteq dom(u)$. There is a
successful reduction $\varphi = \varphi_{2} \ \varphi_{1}$ of $u$,
where $\varphi_{1}$ is a $\{Spr,Sdr\}$-reduction and $\varphi_{2}$
is a $\{Snr\}$-reduction with $dom(\varphi_{2}) = D$ iff
$\pcgr_u|_{D}$ is a tree.
\end{Theorem}
\begin{Proof}
It directly follows from Lemma~\ref{lemma_counting_form} and
Lemma~\ref{lemma_varphi_rem_commute}.
\end{Proof}

Since there are many well known and efficient methods for
determining spanning trees in a graph, it is easy to determine, for
a given set of pointers $D$, whether or not there is a successful
reduction applying string negative rules on exactly the pointers of
$D$ (for a given legal string $u$).

\begin{Example}
We continue the example. By Theorem~\ref{th_snr1} and
Figure~\ref{ex1_subgraph1}, there is a successful reduction $\varphi
= \varphi_{2} \ \varphi_{1}$ of $u$, for some
$\{Spr,Sdr\}$-reduction $\varphi_1$ and $\{Snr\}$-reduction
$\varphi_{2}$ with $dom(\varphi_{2}) = \{2,3,5\}$. Indeed, we can
take for example $\varphi = \textbf{snr}_{5} \ \textbf{snr}_{2} \
\textbf{snr}_{\bar{3}} \ \textbf{spr}_{\bar{7}} \
\textbf{sdr}_{4,6}$.

By Theorem~\ref{th_snr1} and Figure~\ref{ex1_subgraph2}, there is no
successful reduction $\varphi = \varphi_{2} \ \varphi_{1}$ of $u$,
where $\varphi_{1}$ is a $\{Spr,Sdr\}$-reduction and $\varphi_{2}$
is a $\{Snr\}$-reduction with $dom(\varphi_{2}) = \{2,3,4\}$. For
example, $(\textbf{spr}_{5} \ \textbf{spr}_{7})(u) = 6 2 \bar 3 \bar
4 \bar 2 3 4 6$ and thus there is no string pointer rule for pointer
$6$ applicable to this legal string.
\end{Example}

In the next corollary we consider the more general case $|D| \leq
|\zeta| - 1$, instead of $|D| = |\zeta| - 1$ in
Theorem~\ref{th_snr1}.
\begin{Corollary}
Let $u$ be a legal string, and let $D \subseteq dom(u)$. There is a
(successful) reduction $\varphi$ of $u$ such that for all $p \in D$,
$\varphi$ contains either $\textbf{snr}_{p}$ or
$\textbf{snr}_{\bar{p}}$ iff $\pcgr_u|_{D}$ is acyclic.
\end{Corollary}
\begin{Proof}
We first prove the forward implication. By Theorem~\ref{th_snr1},
$\pcgr_u|_{D}$ is a subgraph of a tree, and therefore acyclic.

We now prove the reverse implication. By
Theorem~\ref{pcgr_connected}, $\pcgr_u$ is connected, and since
$\pcgr_u|_{D}$ does not contain cycles, we can add edges $q \in
dom(u) \backslash D$ from $\pcgr_u$ such that the resulting graph is
a tree. Then by Theorem~\ref{th_snr1}, it follows that there is a
(successful) reduction $\varphi$ of $u$ containing either
$\textbf{snr}_{p}$ or $\textbf{snr}_{\bar{p}}$ for all $p \in D$.
\end{Proof}

The previous corollary with $|D| = 1$ shows that the reverse
implication of Corollary~\ref{snr_cor} also holds, since
$\pcgr_u|_{\{p\}}$ acyclic implies that the edge $p$ connects two
different vertices in $\pcgr_u$.
\begin{Theorem} \label{snr_th}
Let $u$ be a legal string and let $p \in dom(u)$. Then
$\textbf{snr}_p$ or $\textbf{snr}_{\bar p}$ is in a (successful)
reduction of $u$ iff $p \in snrdom(u)$.
\end{Theorem}

This theorem can also be proven directly.
\begin{Proof}
To prove the reverse implication, let no reduction of $u$ contain
either $\textbf{snr}_p$ or $\textbf{snr}_{\bar p}$. We prove that $p
\not\in snrdom(u)$. By iteratively applying $\textbf{snr}$,
$\textbf{spr}$ and $\textbf{sdr}$ on pointers that are not equal to
$p$ or $\bar p$, we can reduce $u$ to a legal string $v$ such that
for all $q \in dom(v) \backslash \{p\}$:
\begin{itemize}
\item $qq$ and $\bar q \bar q$ are not substrings of $v$.
\item $q$ is negative in $v$.
\item $q$ does not overlap with any pointer in $dom(v) \backslash
\{p\}$.
\end{itemize}
If $rem_{\{p\}}(v) = \lambda$, then $v$ is equal to either $p \bar
p$, $\bar p p$, $p p$ or $\bar p \bar p$. If $rem_{\{p\}}(v) \not=
\lambda$, then, by the last two conditions, there is a $q \in \Pi$
such that $qq$ is a substring of $rem_{\{p\}}(v)$. Then, by the
first condition, either $q p q$, $q \bar p q$, $q p \bar p q$, $q
\bar p p q$, $q p p q$ or $q \bar p \bar p q$ is a substring of $v$.

Thus, either $q p q$, $q \bar p q$, $p \bar p$, $\bar p p$, $p p$ or
$\bar p \bar p$ is a substring of $v$. Since no reduction of $u$
contains $\textbf{snr}_p$ or $\textbf{snr}_{\bar p}$, the last two
cases are not possible. The first two cases correspond to the
following part of $\redgr{v}$.
$$
\xymatrix @=33pt{ ... \redge[r] &
p  \dedge[r] & p \redge[r] &
q  \dedge[r] & q \redge[r] &
p  \dedge[r] & p \redge[r] &
...  }
$$
The cases where $p \bar p$ or $\bar p p$ is a substring of $v$
correspond to the following part of $\redgr{v}$
$$
\xymatrix @=33pt{ ... \redge[r] &
p  \dedge[r] & p \redge[r] &
p  \dedge[r] & p \redge[r] &
...  }
$$
Consequently, in either case, the two desire edges of $\redgr{v}$
with vertices labelled by $p$ belong to the same connected
component. Thus $p \not\in snrdom(v)$. By
Corollary~\ref{snrdom_cor}, $p \not\in snrdom(u)$.
\end{Proof}

\section{The Order of Loop Recombination} \label{sect_order_snr}
In this section we strengthen Theorem~\ref{th_snr1} to allow one to
determine exactly which orderings of string negative rules in
$\varphi_2$ are possible. First we introduce orderings in a tree,
which is similar to topological orderings in a directed acyclic
graph. Here we order the edges instead of the vertices.

\begin{Definition}
Let $T = (V,E,\epsilon)$ be a tree. An \emph{edge-topological
ordering of $T$ (with root $R \in V$)} is a linear order $E' = (e_1,
e_2, \ldots, e_n)$ on $E$ such that if $\epsilon(e_i) = \{C_x,
C_y\}$, $\epsilon(e_j) = \{C_y, C_z\}$, and $C_y$ ($C_z$, resp.) is
the father of $C_x$ ($C_y$, resp.) in $T$ considering $R$ as the
root of $T$, then $j > i$.
\end{Definition}

\begin{Example}
We continue the example. Consider again tree $\pcgr_u|_{D_1}$ shown
in Figure~\ref{ex1_subgraph1}. Taking $R$ as the root of
$\pcgr_u|_{D_1}$, it follows that $(3,2,5)$ is an edge-topological
ordering of $\pcgr_u|_{D_1}$.
\end{Example}

The next theorem characterizes exactly the possible orders in which
string negative rules that can be applied in a successful reduction
of a given legal string.

\begin{Theorem} \label{snr_theorem2}
Let $u$ be a legal string, and let $L = (p_1, p_2, \ldots, p_n)$ be
an ordered set over $dom(u)$. There is a successful reduction
$$
\varphi = \varphi_{n+1} \ \textbf{snr}_{\tilde{p}_n} \ \varphi_{n} \
\textbf{snr}_{\tilde{p}_{n-1}} \ \cdots \ \varphi_{2} \
\textbf{snr}_{\tilde{p}_1} \ \varphi_{1}
$$
of $u$, for some (possible empty) $\{Spr,Sdr\}$-reductions
$\varphi_{1}, \varphi_{2}, \ldots, \varphi_{n+1}$ and $\tilde{p_i}
\in \{p_i, \bar{p_i}\}$ for $1 \leq i \leq n$ and $n \geq 0$ iff
$\pcgr_u|_{L'}$ is a tree with $L' = \{p_1, p_2, \ldots, p_n\}$ and
$L$ is an edge-topological ordering of $\pcgr_u|_{L'}$ with the
linear component $R$ of $\redgr{u}$ as root.
\end{Theorem}
\begin{Proof}
We first prove the forward implication. Clearly, we can postpone the
application of string negative rules, thus
$\textbf{snr}_{\tilde{p}_n} \ \textbf{snr}_{\tilde{p}_{n-1}} \
\cdots \ \textbf{snr}_{\tilde{p}_1} \ \varphi'$ is also a successful
reduction of $u$, where $\varphi' = \varphi_{n+1} \ \varphi_{n} \
\cdots \ \varphi_{1}$. By Theorem~\ref{th_snr1}, $\pcgr_u|_{L'}$ is
a tree.

We prove that $L$ is an edge-topological ordering of $\pcgr_u|_{L'}$
with root $R$. By Theorem~\ref{th_red_spr_sdr}, $\pcgr_{\varphi'(u)}
\approx \pcgr_u|_{L'}$. If $n > 0$, then
$\textbf{snr}_{\tilde{p}_{1}}$ is applicable to $\varphi'(u)$. By
Theorem~\ref{lemma_cyclic_comp_2vert}, edge $p_1$ is connected to a
leaf of $\pcgr_{\varphi'(u)}$. By Theorem~\ref{th_rf_pc},
$\pcgr_{(\textbf{snr}_{\tilde{p}_{1}} \ \varphi')(u)}$ is isomorphic
to the graph obtained from $\pcgr_{\varphi'(u)}$ by removing $p_1$
and its leaf. Now (assuming $n > 1$), since
$\textbf{snr}_{\tilde{p}_{2}}$ is applicable to
$(\textbf{snr}_{\tilde{p}_{1}} \ \varphi')(u)$, $p_2$ is connected
to a leaf in $\pcgr_{(\textbf{snr}_{\tilde{p}_{1}} \ \varphi')(u)}$.
By iterating this argument, it follows that $L$ is an
edge-topological ordering of $\pcgr_{\varphi'(u)} \approx
\pcgr_u|_{L'}$ with root $R$.

We now prove the reverse implication. Since $\pcgr_u|_{L'}$ is a
tree, by Theorem~\ref{th_snr1} there is a successful reduction
$\varphi = \varphi_{2} \ \varphi_{1}$ of $u$, where $\varphi_{1}$ is
a $\{Spr,Sdr\}$-reduction and $\varphi_{2}$ is a $\{Snr\}$-reduction
with $dom(\varphi_{2}) = L'$. Let $L$ be an edge-topological
ordering of $\pcgr_u|_{L'}$ with the linear component $R$ of
$\redgr{u}$ as root. Again, by Theorem~\ref{th_red_spr_sdr},
$\pcgr_u|_{L'} \approx \pcgr_{\varphi_1(u)}$.

If $n > 0$, then $p_1$ is connected to a leaf $C_1$ of $PC_1$.
Consequently, $C_1$ has only vertices labelled by $p_1$. By
Lemma~\ref{lemma_cyclic_comp_2vert}, $\textbf{snr}_{\tilde{p}_{1}}$
is applicable to $\varphi_{1}(u)$ for some $\tilde{p_1} \in \{p_1,
\bar{p_1}\}$. By Theorem~\ref{th_rf_pc},
$\pcgr_{(\textbf{snr}_{\tilde{p}_{1}} \ \varphi_1)(u)}$ is
isomorphic to the graph obtained from $\pcgr_{\varphi_1(u)}$ by
removing $p_1$ and its leaf. By iterating this argument, it follows
that $\textbf{snr}_{\tilde{p}_n} \ \textbf{snr}_{\tilde{p}_{n-1}} \
\cdots \ \textbf{snr}_{\tilde{p}_1}$ is a successful reduction of
$u$ for some $\tilde{p_i} \in \{p_i, \bar{p_i}\}$ and $1 \leq i \leq
n$ with $n \geq 0$.
\end{Proof}

\begin{Example}
We continue the example. Since $(3,2,5)$ is an edge-topological
ordering of tree $\pcgr_u|_{D_1}$ with root $R$, by
Theorem~\ref{snr_theorem2} there is a successful reduction $\varphi
= \varphi_{2} \ \varphi_{1}$ of $u$, for some
$\{Spr,Sdr\}$-reduction $\varphi_1$ and $\varphi_{2} =
\textbf{snr}_{\tilde{5}} \ \textbf{snr}_{\tilde{2}} \
\textbf{snr}_{\tilde{3}}$ for some $\tilde{p} \in \{p, \bar{p}\}$
for $p \in \{2,3,5\}$. Indeed, we can take for example $\varphi =
\textbf{snr}_{5}$ $\textbf{snr}_{2} \ \textbf{snr}_{\bar{3}} \
\textbf{spr}_{\bar{7}}$ $\textbf{sdr}_{4,6}$.
\end{Example}

We say that two reduction rules $\rho_1$ and $\rho_2$ can be applied
in \emph{parallel} to $u$ if both $\rho_2 \ \rho_1$ and $\rho_1 \
\rho_2$ are applicable to $u$ (see \cite{DBLP:conf/dna/HarjuLPR04}).

\begin{Corollary} \label{snr_parallel1}
Let $u$ be a legal string, and $p,q \subseteq dom(u)$ with $p \not=
q$. Then $\textbf{snr}_{\tilde{p}}$ and $\textbf{snr}_{\tilde q}$
can be applied in parallel to $u$ for some $\tilde{p} \in \{p, \bar
p\}$, $\tilde{q} \in \{q, \bar q\}$ iff there is a spanning tree $T$
in $\pcgr_u$ such that $p$ and $q$ both connect to leaves
(considering the linear component of $\redgr{u}$ as the root).
\end{Corollary}

The next corollary considers the case whether or not
$\textbf{snr}_{\tilde{p}}$ and $\textbf{snr}_{\tilde q}$ can
\emph{eventually} be applied in parallel.
\begin{Corollary} \label{snr_parallel2}
Let $u$ be a legal string, and $p,q \subseteq dom(u)$ with $p \not=
q$. Then $\textbf{snr}_{\tilde{p}}$ and $\textbf{snr}_{\tilde q}$
can be applied in parallel to $\varphi(u)$ for some $\tilde{p} \in
\{p, \bar p\}$, $\tilde{q} \in \{q, \bar q\}$ and reduction
$\varphi$ iff there is a spanning tree $T$ in $\pcgr_u$ such that
there is no simple walk in $T$ from the linear component of
$\redgr{u}$ (the root) to another vertex of $T$ containing both
edges $p$ and $q$.
\end{Corollary}

\begin{figure}
$$
\xymatrix{
& R \ar@{-}[d]^6 \\
& C_2 \ar@{-}[dl]_2 \ar@{-}[dr]^4 \\
C_1 & & C_3
}
$$
\caption{A subgraph of the pointer-component graph from the
Example.} \label{ex1_subgraph3}
\end{figure}
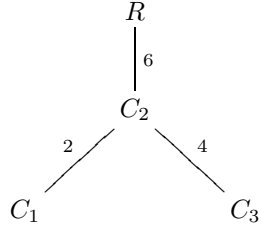

\begin{Example}
We continue the example. Let $D_3 = \{2,4,6\}$. Then in tree
$\pcgr_u|_{D_3}$, depicted in Figure~\ref{ex1_subgraph3}, there is
no simple walk from $R$ to another vertex of $\pcgr_u|_{D_3}$
containing both edges $2$ and $4$. By Corollary~\ref{snr_parallel2},
$\textbf{snr}_{\tilde{2}}$ and $\textbf{snr}_{\tilde 4}$ can be
applied in parallel to $\varphi(u)$ for some $\tilde{2} \in \{2,
\bar 2\}$, $\tilde{4} \in \{4, \bar 4\}$ and reduction $\varphi$ of
$u$. Indeed, if we take $\varphi = \textbf{spr}_{\bar 7} \
\textbf{sdr}_{3,5}$, then $\textbf{snr}_{2}$ and $\textbf{snr}_{4}$
can be applied in parallel to $\varphi(u) = 6 2 2 4 4 6$.
\end{Example}

\section{Conclusion} \label{sect_conclusion}
This paper showed that one can efficiently determine the possible
sequences of loop recombination operations that can be applied in
the transformation of a given gene from its micronuclear to its
macronuclear form. Formally, one can determine which string negative
rules can be applied in which order to a legal string $u$, given
only the reduction graph of $u$. This is characterized in terms of
graphs defined on the reduction graphs. Future research could focus
on similar characterizations for the string positive rules and the
string double rules.

\section*{Acknowledgments}
This research was supported by the Netherlands Organization for
Scientific Research (NWO) project 635.100.006 `VIEWS'.

\bibliography{../gene_assembly}

\end{document}